\documentstyle[twoside,psfig,cite,amssymb]{article}

\catcode`\@=11
\long\def\@makefntext#1{
\protect\noindent \hbox to 3.2pt {\hskip-.9pt  
$^{{\eightrm\@thefnmark}}$\hfil}#1\hfill}		

\def\@makefnmark{\hbox to 0pt{$^{\@thefnmark}$\hss}}	
	
\def\ps@myheadings{\let\@mkboth\@gobbletwo
\def\@oddhead{\hbox{}
\rightmark\hfil\eightrm\thepage}   
\def\@oddfoot{}\def\@evenhead{\eightrm\thepage\hfil
\leftmark\hbox{}}\def\@evenfoot{}
\def\sectionmark##1{}\def\subsectionmark##1{}}

\oddsidemargin=\evensidemargin
\newcounter{sectionc}\newcounter{subsectionc}\newcounter{subsubsectionc}
\renewcommand{\section}[1] {\vspace{12pt}\addtocounter{sectionc}{1} 
\setcounter{subsectionc}{0}\setcounter{subsubsectionc}{0}\noindent 
	{\tenbf\thesectionc. #1}\par\vspace{5pt}}
\renewcommand{\subsection}[1] {\vspace{12pt}\addtocounter{subsectionc}{1} 
	\setcounter{subsubsectionc}{0}\noindent 
	{\bf\thesectionc.\thesubsectionc. {\kern1pt \bfit #1}}\par\vspace{5pt}}
\renewcommand{\subsubsection}[1] {\vspace{12pt}\addtocounter{subsubsectionc}{1}
	\noindent{\tenrm\thesectionc.\thesubsectionc.\thesubsubsectionc.
	{\kern1pt \tenit #1}}\par\vspace{5pt}}
\newcommand{\nonumsection}[1] {\vspace{12pt}\noindent{\tenbf #1}
	\par\vspace{5pt}}

\newcounter{appendixc}
\newcounter{subappendixc}[appendixc]
\newcounter{subsubappendixc}[subappendixc]
\renewcommand{\thesubappendixc}{\Alph{appendixc}.\arabic{subappendixc}}
\renewcommand{\thesubsubappendixc}
	{\Alph{appendixc}.\arabic{subappendixc}.\arabic{subsubappendixc}}

\renewcommand{\appendix}[1] {\vspace{12pt}
        \refstepcounter{appendixc}
        \setcounter{figure}{0}
        \setcounter{table}{0}
        \setcounter{lemma}{0}
        \setcounter{theorem}{0}
        \setcounter{corollary}{0}
        \setcounter{definition}{0}
        \setcounter{equation}{0}
        \renewcommand{\thefigure}{\Alph{appendixc}.\arabic{figure}}
        \renewcommand{\thetable}{\Alph{appendixc}.\arabic{table}}
        \renewcommand{\theappendixc}{\Alph{appendixc}}
        \renewcommand{\thelemma}{\Alph{appendixc}.\arabic{lemma}}
        \renewcommand{\thetheorem}{\Alph{appendixc}.\arabic{theorem}}
        \renewcommand{\thedefinition}{\Alph{appendixc}.\arabic{definition}}
        \renewcommand{\thecorollary}{\Alph{appendixc}.\arabic{corollary}}
        \renewcommand{\theequation}{\Alph{appendixc}.\arabic{equation}}
        \noindent{\tenbf Appendix \theappendixc #1}\par\vspace{5pt}}
\newcommand{\subappendix}[1] {\vspace{12pt}
        \refstepcounter{subappendixc}
        \noindent{\bf Appendix \thesubappendixc. {\kern1pt \bfit #1}}
	\par\vspace{5pt}}
\newcommand{\subsubappendix}[1] {\vspace{12pt}
        \refstepcounter{subsubappendixc}
        \noindent{\rm Appendix \thesubsubappendixc. {\kern1pt \tenit #1}}
	\par\vspace{5pt}}

\newcommand{\textlineskip}{\baselineskip=13pt}
\newcommand{\smalllineskip}{\baselineskip=10pt}

\def\eightcirc{
\begin{picture}(0,0)
\put(4.4,1.8){\circle{6.5}}
\end{picture}}
\def\eightcopyright{\eightcirc\kern2.7pt\hbox{\eightrm c}} 

\newcommand{\copyrightheading}[1]
	{\vspace*{-2.5cm}\smalllineskip{\flushleft
	{\footnotesize Modern Physics Letters A, #1}\\
	{\footnotesize $\eightcopyright$\, World Scientific Publishing
	 Company}\\
	 }}

\def\abstracts#1#2#3{{
	\centering{\begin{minipage}{4.5in}\footnotesize\baselineskip=10pt
	\parindent=0pt #1\par 
	\parindent=15pt #2\par
	\parindent=15pt #3
	\end{minipage}}\par}} 

\newcommand{\bibit}{\nineit}
\newcommand{\bibbf}{\ninebf}
\renewenvironment{thebibliography}[1]
	{\frenchspacing
	 \ninerm\baselineskip=11pt
	 \begin{list}{\arabic{enumi}.}
        {\usecounter{enumi}\setlength{\parsep}{0pt}     
	 \setlength{\leftmargin 12.7pt}{\rightmargin 0pt} 
         \setlength{\itemsep}{0pt} \settowidth
	{\labelwidth}{#1.}\sloppy}}{\end{list}}

\newcounter{itemlistc}
\newcounter{romanlistc}
\newcounter{alphlistc}
\newcounter{arabiclistc}
\newenvironment{itemlist}
    	{\setcounter{itemlistc}{0}
	 \begin{list}{$\bullet$}
	{\usecounter{itemlistc}
	 \setlength{\parsep}{0pt}
	 \setlength{\itemsep}{0pt}}}{\end{list}}

\newenvironment{romanlist}
	{\setcounter{romanlistc}{0}
	 \begin{list}{$($\roman{romanlistc}$)$}
	{\usecounter{romanlistc}
	 \setlength{\parsep}{0pt}
	 \setlength{\itemsep}{0pt}}}{\end{list}}

\newenvironment{alphlist}
	{\setcounter{alphlistc}{0}
	 \begin{list}{$($\alph{alphlistc}$)$}
	{\usecounter{alphlistc}
	 \setlength{\parsep}{0pt}
	 \setlength{\itemsep}{0pt}}}{\end{list}}

\newcommand{\fcaption}[1]{
        \refstepcounter{figure}
        \setbox\@tempboxa = \hbox{\footnotesize Fig.~\thefigure. #1}
        \ifdim \wd\@tempboxa > 5in
           {\begin{center}
        \parbox{5in}{\footnotesize\smalllineskip Fig.~\thefigure. #1}
            \end{center}}
        \else
             {\begin{center}
             {\footnotesize Fig.~\thefigure. #1}
              \end{center}}
        \fi}

\newcommand{\tcaption}[1]{
        \refstepcounter{table}
        \setbox\@tempboxa = \hbox{\footnotesize Table~\thetable. #1}
        \ifdim \wd\@tempboxa > 5in
           {\begin{center}
        \parbox{5in}{\footnotesize\smalllineskip Table~\thetable. #1}
           \end{center}}
        \else
             {\begin{center}
             {\footnotesize Table~\thetable. #1}
              \end{center}}
        \fi}

\def\@citex[#1]#2{\if@filesw\immediate\write\@auxout
	{\string\citation{#2}}\fi
\def\@citea{}\@cite{\@for\@citeb:=#2\do
	{\@citea\def\@citea{,}\@ifundefined
	{b@\@citeb}{{\bf ?}\@warning
	{Citation `\@citeb' on page \thepage \space undefined}}
	{\csname b@\@citeb\endcsname}}}{#1}}

\newif\if@cghi
\def\cite{\@cghitrue\@ifnextchar [{\@tempswatrue
	\@citex}{\@tempswafalse\@citex[]}}
\def\citelow{\@cghifalse\@ifnextchar [{\@tempswatrue
	\@citex}{\@tempswafalse\@citex[]}}
\def\@cite#1#2{{$\null^{#1}$\if@tempswa\typeout
	{IJCGA warning: optional citation argument 
	ignored: `#2'} \fi}}

\def\pmb#1{\setbox0=\hbox{#1}
	\kern-.025em\copy0\kern-\wd0
	\kern.05em\copy0\kern-\wd0
	\kern-.025em\raise.0433em\box0}

\def\fnt#1#2{\footnotetext{\kern-.3em
	{$^{\mbox{\scriptsize #1}}$}{#2}}}

\def\fpage#1{\begingroup
\voffset=.3in
\thispagestyle{empty}\begin{table}[b]\centerline{\footnotesize #1}
	\end{table}\endgroup}

\def\runninghead#1#2{\pagestyle{myheadings}
\markboth{{\protect\footnotesize\it{\quad #1}}\hfill}
{\hfill{\protect\footnotesize\it{#2\quad}}}}
\font\tenrm=cmr10
\font\tenit=cmti10 
\font\tenbf=cmbx10
\font\bfit=cmbxti10 at 10pt
\font\ninerm=cmr9
\font\nineit=cmti9
\font\ninebf=cmbx9
\font\eightrm=cmr8

\def\qed{\hbox{${\vcenter{\vbox{			
   \hrule height 0.4pt\hbox{\vrule width 0.4pt height 6pt
   \kern5pt\vrule width 0.4pt}\hrule height 0.4pt}}}$}}

\begin{document}
\setlength{\textheight}{7.7truein}  

\runninghead{Regularized Green's Function/Inverse Square Potential
}{Regularized Green's Function/Inverse Square Potential
}

\normalsize\textlineskip
\thispagestyle{empty}
\setcounter{page}{1}

\copyrightheading{}			

\vspace*{0.88truein}

\fpage{1}
\centerline{\bf REGULARIZED GREEN'S FUNCTION} 
\baselineskip=13pt
\centerline{\bf FOR THE INVERSE SQUARE POTENTIAL}
\vspace*{0.37truein}
\centerline{\footnotesize HORACIO E. CAMBLONG\footnote{E-mail: camblong@usfca.edu}}
\baselineskip=12pt
\centerline{\footnotesize\it 
Department of Physics, University of San Francisco, San Francisco, CA 94117-1080}
\vspace*{10pt}

\centerline{\footnotesize CARLOS R. ORD\'{O}\~{N}EZ\footnote{E-mail: ordonez@uh.edu}}
\baselineskip=12pt
\centerline{\footnotesize\it Department of Physics, University of Houston, Houston, TX 77204-5506}
\baselineskip=10pt
\centerline{\footnotesize\it 
World Laboratory 
Center for Pan-American Collaboration in Science and
Technology,}
\baselineskip=10pt
\centerline{\footnotesize\it  University of Houston Center, Houston, TX 77204-5506}
\vspace*{0.225truein}


\vspace*{0.21truein}
\abstracts{A Green's function approach is presented for 
the $D$-dimensional inverse square 
potential in quantum mechanics.
This approach 
is implemented by the introduction of hyperspherical coordinates and 
the use of a real-space regulator in the regularized version of the model.
The application of Sturm-Liouville theory
yields a closed expression for the 
radial energy Green's function.
Finally,
the equivalence with a recent path-integral treatment of the same
problem is explicitly shown.}{}{}

\vspace*{1pt}\textlineskip      
\section{Introduction}    
\vspace*{-0.5pt}
\noindent
The use of path integrals for the analysis of the bound-state and
scattering sectors in quantum mechanics has significantly advanced in the 
past three decades.\cite{schulman,kleinert,grosche} However,
much work remains to be done, a situation that 
is particularly evident for the case of singular 
potentials.\cite{fra:71} Among these, the
contact interactions\cite{albeverio,tho:79,hua:92}
and the inverse square potential\cite{isp_classics1,isp_classics2} 
stand out;
these problems have
 been the subject of active research in the past few years,
mainly through the direct use of the Schr\"{o}dinger equation combined with
appropriate regularization and renormalization analyses
{\em \`{a} la\/} field 
theory.\cite{jackiw:delta,delta_generic,gup:93,cam:isp,cam:dtI,cam:dtII,Beane}
Recently we have started a 
program centered on the use of path
integrals for a systematic study of 
bound states and 
of the connection between field theory and
quantum mechanics.\cite{cam:pi_bs,cam:delta}
We expect that our approach may shed light on the formidable
problem posed by bound states in quantum field theory.\cite{wei:95}
In addition to introducing a technique of infinite summations 
of perturbation theory, we 
have specifically found analytic solutions for the
two most outstanding singular interactions:
the inverse square 
potential\cite{cam:pi_bs}
and the two-dimensional delta-function 
interaction.\cite{cam:delta}

In this paper we now investigate the quantum-mechanical
properties of the inverse square potential from a
complementary viewpoint, using an intrinsically 
 {\em nonperturbative\/} operator approach. 
This is accomplished by directly considering the problem 
associated with the 
energy Green's function $G(E)$, defined as the
Fourier transform of the path-integral
propagator $K(T)$ [multiplied by the Heaviside function $\theta (T)$].
The Green's function is then computed
by solving a Sturm-Liouville problem in real 
space.\cite{morse_feshbach,arfken,stakgold}
Our final results are in complete agreement 
with the ones we obtained earlier by means of a
path-integral perturbative approach.\cite{cam:pi_bs}
More precisely, we 
explicitly show that both approaches reproduce identical
energy Green's functions.

The relevance of this computation is 
highlighted by its recent application to a simple realization of a quantum anomaly
in Nature: 
the interaction of an electron with a polar molecule.\cite{cam:dipole}
As shown in 
Ref.~\citen{cam:dipole}, the scale symmetry properties of the inverse square potential
are inherited by the full-fledged dipole
potential describing the physics of this system---electron binding
occurs for a sufficiently strong coupling (dipole moment)
and this amounts to quantum mechanical symmetry breaking.
Our calculations in this paper provide additional 
support for this remarkable and unusual result.
This mechanism is further analyzed in Sec.~4.   

\section{General Framework}
\label{sec:general_framework}
\noindent
The path-integral treatment for a 
$D$-dimensional 
nonrelativistic particle 
is based on the quantum-mechanical propagator
\begin{equation}
K_{D}({\bf r''}, {\bf r'}; t^{''},t^{'}) 
= 
\lim_{N \rightarrow \infty}
\left( \frac{M}{2 \pi i \epsilon \hbar} \right)^{DN/2}
\,
\left[
\prod_{k=1}^{N-1} \int_{ \mathbb{R}^{D} } d^{D} {\bf r}_{k} \right]
\;
e^{i S^{(N)}/\hbar}
\;  .
\label{eq:propagator_QM}
\end{equation}
Equation~(\ref{eq:propagator_QM})
has been explicitly formulated in Cartesian coordinates
by means of
a time lattice $t_{j}= t'+j \epsilon$
 [where
$\epsilon=
(t''-t')/N$, with $j=0, \cdots,N$,
while $t_{0} \equiv t'$ and $t_{N} \equiv t''$],
such that $ {\bf r}_{j} = {\bf r} (t_{j})$,
with the end points being
${\bf r}_{0} \equiv {\bf r'}$
and 
${\bf r}_{N} \equiv {\bf r''}$.
The corresponding discrete action 
in Eq.~(\ref{eq:propagator_QM}) 
is  $S^{(N)}=   \sum_{j=0}^{N-1} S^{(N)}_{j}$, with
$S^{(N)}_{j}
= M
( {\bf r}_{j+1}
- {\bf r}_{j}  )^{2}/2\epsilon -
\epsilon V({\bf r}_{j}, t_{j})
 $, for a particle of mass $M$ subject to a 
potential $V({\bf r},t)$.
The connection with the operator approach is established by means of
\begin{equation} 
K_{D}({\bf r''}, {\bf r'}; t^{''},t^{'}) 
=
\left\langle {\bf r''}
 \left| 
\hat{T}
\exp \left[
 -\frac{i}{\hbar} \int_{t'}^{t''}
\hat{H} dt 
\right]
\right| {\bf r'}
\right\rangle   
\; ,
\label{eq:position_propagator2}
\end{equation}
where $\hat{T}$ is the time-ordering operator
and  $\hat{H}$ is the Hamiltonian.
For a time-independent Hamiltonian, the dependence of
Eq.~(\ref{eq:position_propagator2}) 
with respect to the times
$t'$ and $t''$
is only through
the time difference $T=t''-t'$. For this all-important
case, the energy Green's function is then defined
as a Fourier transform of the Green's function
obtained from
(\ref{eq:propagator_QM})-(\ref{eq:position_propagator2})
by multiplication with the Heaviside function $\theta (T)$,
i.e., 
\begin{equation}
G_{D} ({\bf r''},{\bf r'};E) 
=
\frac{1}{i\hbar}
\int_{0}^{\infty} dT  e^{iET/\hbar}
K_{D}({\bf r''}, {\bf r'}; t'',t')
\; .
\label{eq:GF_as_FT}
\end{equation}
Equations~(\ref{eq:position_propagator2}) and (\ref{eq:GF_as_FT}) 
establish the operator form of the energy Green's function
in its configuration-space representation,
which explicitly becomes
\begin{equation}
G_{D} ({\bf r''},{\bf r'};E) =
\left\langle {\bf r''}
\left|
\left( 
E - \hat{H} + i \epsilon 
\right)^{-1}
 \right|
{\bf r'}
\right\rangle   
\; ,
\label{eq:GF_as_operator}
\end{equation}
where $i\epsilon=i0^{+}$ is a small imaginary part.
The operator structure of Eq.~(\ref{eq:GF_as_operator}) 
implies the differential equation
\begin{equation}
\left\{
\nabla^{2}_{\bf r'} + \frac{2M}{ \hbar^{2} } 
\left[ E - V({\bf r'}) 
\right]
\right\}
{\mathcal G}_{D}({\bf r''}, {\bf r'}; E) 
=
\delta ({\bf r''}-{\bf r'})
\;  ,
\label{eq:GF_diff_eq}
\end{equation} 
in which
we have conveniently defined the rescaled Green's function
\begin{equation}
{\mathcal G}_{D} 
 ({\bf r''},{\bf r'} ; E) 
=
  \frac{\hbar^{2}}{2M} \,
G_{D} 
 ({\bf r''},{\bf r'} ; E) 
\;  .
\label{eq:rescaled_GF}
\end{equation}

For the particular case of a central potential,
Eq.~(\ref{eq:GF_diff_eq})
can be solved by separation of variables. This procedure can be
systematically implemented by considering
a complete set of angular functions, the
$D$-dimensional hyperspherical harmonics
$Y_{l m} ({\bf \Omega})$, for which $l$ is the angular momentum 
quantum number and
$m=0,\ldots, d_{l}$, with
$d_{l}= (2l+D-2)(l+D-3)!/l!(D-2)!.$\cite{erd:53}
Then, the 
partial-wave expansion
\begin{equation}
G_{D}({\bf r''}, {\bf r'}; E) 
=
\left( r'' r' \right)^{-(D-1)/2}
\,
\sum_{l= 0}^{\infty}
\sum_{m=1}^{d_{l}}
Y_{l m} ({\bf \Omega''})
Y_{l m}^{ *} ({\bf \Omega'})
G_{l +\nu}(r'',r';E)
\;  
\label{eq:propagator_partial_wave_exp}
\end{equation}
implicitly defines the radial energy Green's function
$
G_{l+\nu}(r'',r';E)
$
for each angular momentum channel $l$
and dimensionality $D= 2(\nu +1)$.
Equation~(\ref{eq:propagator_partial_wave_exp})
has been written to 
explicitly display the property of
interdimensional dependence:\cite{interdimensional}
except for the scale prefactors 
$\left( r'' r' \right)^{-(D-1)/2}$,
the only dependence of 
$G_{D}({\bf r''}, {\bf r'}; E) $
upon the dimensionality
$D$ is through the combination $l+\nu$,
and this information is conveyed by 
the function $G_{l +\nu}(r'',r';E)$.
Likewise
we define a rescaled radial Green's function
${\mathcal G}_{l+\nu}(r'',r';E)$
from the expansion of
${\mathcal G}_{D} 
 ({\bf r''},{\bf r'} ; E) $, just as in 
(\ref{eq:propagator_partial_wave_exp})---or alternatively, 
by directly enforcing the analogue of Eq.~(\ref{eq:rescaled_GF}).
Then,
substitution of Eq.~(\ref{eq:propagator_partial_wave_exp}) in
(\ref{eq:GF_diff_eq}) yields
the radial differential equation
\begin{equation}
\left\{ \frac{d^{2}}{dr'^{2}} + 
\frac{2M}{ \hbar^{2} }
\left[
E - V( r')
\right]
-
\frac{ \left( l + \nu \right)^{2}  
- 1/4}{r'^{2}}
\right\}  
{\mathcal G}_{l+\nu}(r'',r';E)
=
\delta ( r''- r')
\;  ,
\label{eq:GF_radial}
\end{equation}
which stands for  
a particular case of the Sturm-Liouville Green's function 
problem\cite{morse_feshbach,arfken,stakgold}
\begin{equation}
\left\{ \frac{d}{dr'} 
\left[ p(r') 
\frac{d}{dr'} 
\right]
+\eta +
q(r')
\right\}
{\mathcal G} (r'',r';E)
=
\delta ( r''- r')
\;  ,
\label{eq:Sturm_Liouville}
\end{equation}
with $p(r) \equiv 1$,
$\eta= 2ME/\hbar^{2}$, and $q(r)= - 2MV_{\rm eff}(r)/\hbar^{2}$;
here $V_{\rm eff}(r)$ includes, in addition to the potential $V(r)$, the
centrifugal barrier,\cite{centrifugal} 
displayed as the last term on the right-hand side
of Eq.~(\ref{eq:GF_radial}).

Equation~(\ref{eq:Sturm_Liouville})
can be solved by applying the following standard 
technique. Let
$u^{(<)}_{l,\nu} (r)$
and 
$u^{(>)}_{l,\nu} (r)$ 
be the particular
solutions to the corresponding homogeneous differential equation
\begin{equation}
\left\{ \frac{d}{dr} 
\left[ p(r) 
\frac{d}{dr} 
\right]
+\eta +
q(r)
\right\}
u_{l,\nu}(r) = 0
\;  ,
\label{eq:GF_radial_homogeneous}
\end{equation}
subject to the appropriate boundary conditions: 
$u^{(<)}_{l,\nu} (r)$ at the left boundary and 
$u^{(>)}_{l,\nu} (r)$ at the right boundary.
Then,
the Green's function is given by
\begin{equation}
{\mathcal G}_{l +\nu}(r'',r';E)
=
\frac{ u^{(<)}_{l,\nu} 
(r_{<}) u^{(>)}_{l,\nu}  (r_{>}) }{ 
p (r') \;  
W \left\{ 
u^{(<)}_{l,\nu} ,u^{(>)}_{l,\nu} 
\right\} 
(r') }
\;  ,
\label{eq:GF_radial_explicit}
\end{equation}
where 
 $r_{<}$ ($r_{>}$) is the smaller
(larger) of $r'$ and $r''$,
while $W \left\{ u^{(<)}_{l,\nu} ,u^{(>)}_{l,\nu} \right\} (r)$
is the Wronskian of 
$u^{(<)}_{l,\nu} (r)$
and 
$u^{(>)}_{l,\nu} (r)$.\cite{Wronskian}

Solving Eq.~(\ref{eq:GF_radial}) 
requires specific knowledge of the potential
and has to be dealt with on a case-by-case basis.
We now turn our attention to the advertised problem:
the inverse square potential.

\section{Green's Function for the Inverse Square Potential}
\noindent
For the  case of an inverse square potential,
we conveniently write 
\begin{equation}
V ( r ) = -\frac{ \hbar^{2} }{ 2M} \, \frac{\lambda}{r^{2}}
\;  ,
\end{equation} 
so that
\begin{equation}
\left[
\frac{d^{2}}{dr'^{2}} 
 + \eta  -
\frac{ \left( l + \nu \right)^{2}  - \lambda
- 1/4}{r'^{2}}
\right]  
{\mathcal G}_{l+\nu}(r'',r';E)
=
\delta (r''-r')
\;  .
\label{eq:GF_radial_ISP}
\end{equation}
As a particular case of the 
Sturm-Liouville problem~(\ref{eq:Sturm_Liouville}),
Eq.~(\ref{eq:GF_radial_ISP})
can be solved by introducing 
the functions
$u^{(<)}_{l,\nu} (r)$
and 
$u^{(>)}_{l,\nu} (r)$ 
that satisfy the corresponding homogeneous differential equation
\begin{equation}
\left[ \frac{d^{2}}{dr^{2}} + \eta -
\frac{ \left( l + \nu \right)^{2}  - \lambda
- 1/4}{r^{2}}
\right]  
u_{l,\nu}(r) = 0
\;  ,
\label{eq:GF_radial_homogeneous_ISP}
\end{equation}
along with the boundary conditions
outlined in Sec.~2. 
Equation~(\ref{eq:GF_radial_homogeneous_ISP})
is immediately recognized to have solutions 
of the generic form\cite{abr:72}
\begin{equation}
  u_{l,\nu}(r)
=
\sqrt{r}
\,
Z_{s_{l}}
\left(\sqrt{\eta} \, r
\right)
\; ,
\label{eq:ISP_solutions}
\end{equation}
where $Z_{s_{l}} (z) $ represents an appropriate linear combination
of Bessel functions of order
\begin{equation}
s_{l}
=
\sqrt{  \lambda_{l}^{(\ast)}
- \lambda}
\; ,
\label{eq:ISP_order}
\end{equation}
with
\begin{equation}
\lambda_{l}^{(\ast)}
= \left( l +  \nu  \right)^{2}
\;  .
\label{eq:ISP_critical_coupling}
\end{equation}
Simple inspection of
Eqs.~(\ref{eq:GF_radial_ISP}) and (\ref{eq:GF_radial_homogeneous_ISP})
shows that the only effect of adding an inverse square potential 
$V ( r ) = -\hbar^{2} \lambda/2Mr^{2}$ 
to a free particle
is a shift in the angular momentum quantum number,
\begin{equation}
l+ \nu \rightarrow s_{l}
=\sqrt{ (l+\nu)^{2} - \lambda}
\; .
\label{eq:ISP_order_replacement_from_free_part}
\end{equation}
This analysis shows that 
$\lambda_{l}^{(\ast)} $ in Eq.~(\ref{eq:ISP_critical_coupling})
plays the role of a critical coupling, i.e.,
the nature of the solutions changes abruptly around the value
$ \lambda =\lambda_{l}^{(\ast)} $, for
any state with angular momentum $l$.\cite{cam:dtII} 
Thus,
$\lambda_{l}^{(\ast)} $
represents the threshold separating the two coupling-strength
regimes:
\begin{itemlist}
\item
 subcritical or weak coupling,
$ \lambda <\lambda_{l}^{(\ast)} $ (i.e., real $s_{l} $);
\item
 supercritical or strong coupling,
$ \lambda >\lambda_{l}^{(\ast)} $ (i.e., imaginary $s_{l} $).
\end{itemlist}
In the latter case the order can be rewritten as
 $s_{l} =i \Theta_{l}$, 
with a real parameter
\begin{equation}
\Theta_{l}
= \sqrt{ \lambda - \lambda_{l}^{(\ast)} }
\;  .
\label{eq:Theta}
\end{equation}
Classification of the
nature of the solutions can be fully accomplished by also considering
the values of the  energy $E=\hbar^{2}\eta/2M$:
\begin{romanlist}
 \item  scattering states amount to 
$\eta=k^{2}>0$, for which the Bessel functions
in Eq.~(\ref{eq:ISP_solutions}) have real argument $kr $;
\item
 bound states  amount to
$\eta=-\kappa^{2} < 0$, 
for which the Bessel functions
have imaginary argument $kr = i \kappa r $.
\end{romanlist}
As a result, the solutions~(\ref{eq:ISP_solutions})
fall into one of the following four families:
\begin{romanlist}
\item
subcritical bound-state sector:
\begin{displaymath}
u_{l,\nu}(r)/\sqrt{r}
=
\mbox{\boldmath\large  $\left\{  \right.$ } \! \! \!
I_{s_{l}}(\kappa r)
\mbox{\boldmath\large  $,$ } \!   \!
K_{s_{l}} (\kappa r)
\mbox{\boldmath\large  $\left.  \right\}$ } \! \!
\; ;
\end{displaymath}
\item
subcritical scattering sector
\begin{displaymath}
u_{l,\nu}(r)/\sqrt{r}
 =
\mbox{\boldmath\large  $\left\{  \right.$ } \! \! \!
H^{(1)}_{s_{l}}(kr)
\mbox{\boldmath\large  $,$ }  \!   \!
H^{(2)}_{s_{l}}(kr)
\mbox{\boldmath\large  $\left.  \right\}$ } \! \!
\; ;
\end{displaymath}
\item
supercritical
bound-state sector:
\begin{displaymath}
u_{l,\nu}(r)/\sqrt{r}
=
\mbox{\boldmath\large  $\left\{  \right.$ } \! \! \!
I_{i\Theta_{l}} (\kappa r)
\mbox{\boldmath\large  $,$ } \!   \!
K_{i\Theta_{l}} (\kappa r)
\mbox{\boldmath\large  $\left.  \right\}$ } \! \!
\;  ;
\end{displaymath}
\item
supercritical
scattering sector:
\begin{displaymath}
u_{l,\nu}(r)/\sqrt{r}
=
\mbox{\boldmath\large  $\left\{  \right.$ } \! \! \!
H^{(1)}_{i\Theta_{l}}(kr)
\mbox{\boldmath\large  $,$ } \!   \!
H^{(2)}_{i\Theta_{l}}(kr)
\mbox{\boldmath\large  $\left.  \right\}$ } \! \!
\;  .
\end{displaymath}
\end{romanlist}
Here the 
 symbol
$\mbox{\boldmath\large  $\left\{  \right.$ } \! \! \!
\mbox{\boldmath\large  $,$ } \! \! \!
\mbox{\boldmath\large  $\left.  \right\}$ } \! \!
$
represents a linear combination and
$H_{s_{l}}^{(1,2)} (z)$ stand for the Hankel functions
of the first and second kinds, whereas
 $I_{s_{l}}(z)$ and $K_{s_{l}}(z)$
stand for the modified Bessel functions of the first and second kinds,
respectively.\cite{abr:72}

The analysis of Ref.~\citen{cam:dtII} applies almost verbatim to the 
Green's function formulation,
with the result that regularization
is only needed in the supercritical coupling regime.
However,
in this paper---just as in Refs.~\citen{gup:93} and~\citen{cam:isp}---we regularize
the inverse square potential by
introducing a real-space regulator $a$,
a procedure that is different from 
the dimensional regularization
of Ref.~\citen{cam:dtII}.
The introduction of a regulator
is necessary for strong coupling because
the inverse square potential fails to provide a
discriminating boundary condition at the 
origin.\cite{cam:dtII} 
Thus, this singular behavior  
leads to the emergence of a scale---a behavior known as dimensional 
transmutation.\cite{col:73}
In some sense, the key to a successful
regularization procedure
is the restoration of a sensible boundary condition
at the origin.
In particular, a real-space regulator allows one to restore the usual
homogeneous boundary condition, but now displaced away from the origin
to a radial position $r=a$.
In short, 
 the boundary conditions
satisfied by the Green's function are:
\begin{itemlist}
\item
boundary condition at 
infinity,
\begin{equation}   
u^{(>)}_{l,\nu} (r) 
 \stackrel{(r \rightarrow \infty)}{\longrightarrow} 0
\; ;
\label{eq:BC_infty}
\end{equation}
\item
boundary condition at $r=a$,
\begin{equation}   
u^{(<)}_{l,\nu} (a) = 0 
\; .
\label{eq:BC_a}
\end{equation}
\end{itemlist}
In what follows and as anticipated for the supercritical regime, 
$s_{l} =i \Theta_{l}$, 
with $\Theta_{l}$ 
real and given by Eq.~(\ref{eq:Theta}).
From the behavior of Bessel functions\cite{wat:44}
and the boundary conditions,
one concludes that, for the bound-state sector of the theory,
\begin{equation}
u^{(<)}_{l,\nu} (r)
\propto
\sqrt{r}
\,
\left[
 K_{s_{l}}(\kappa a) I_{s_{l}}(\kappa r)
-
I_{s_{l}}(\kappa a) K_{s_{l}}(\kappa r)
\right]
\;  
\end{equation}
[which satisfies 
condition~(\ref{eq:BC_a})]
and
\begin{equation}
u^{(>)}_{l,\nu} (r)
 \propto 
\sqrt{r}
\,
 K_{s_{l}}(\kappa r)
\;  
\end{equation}
[which satisfies 
condition~(\ref{eq:BC_infty})].
In addition, from the well-known properties of the Sturm-Liouville
equation, the Wronskian of any two Bessel functions $Z(z)$ 
is proportional to $1/z$; for the problem at hand, it suffices 
to know (for example, from the small-argument behavior of the Bessel
functions) that
$W \left\{  K_{s}(z), I_{s}(z) \right\}=1/z$.\cite{abr:72}
Then, 
Eq.~(\ref{eq:GF_radial}) yields straightforwardly 
the desired regularized radial energy Green's function
\begin{eqnarray}
G_{l+\nu} 
\left(
\left. 
r'', r' ;  
E
\right| 
\lambda; a
\right)
& =  &
- \frac{2M}{\hbar^{2}} \sqrt{r'r''}
\; 
\frac{ K_{i\Theta_{l}} (\kappa r_{>})}{ K_{i\Theta_{l}}(\kappa a)}
\nonumber
\\
& \times &
\left[
K_{i\Theta_{l}}(\kappa a) I_{i\Theta_{l}}(\kappa r_{<})
-
I_{i\Theta_{l}}(\kappa a) K_{i\Theta_{l}}(\kappa r_{<})
\right]
\;  ,
\label{eq:GreenF_l_reg}
\end{eqnarray}
with $\kappa=
\sqrt{-2ME}/\hbar$.

Analysis of the scattering sector 
requires an analytic continuation to positive energies, with a replacement of
 the Bessel functions in Eq.~(\ref{eq:GreenF_l_reg})
as follows:
$I_{s} (\kappa r) 
\rightarrow  
(-i)^{s}
J_{s} (k r) $
and
$K_{s} (\kappa r) 
\rightarrow 
\pi i^{s+1}H^{(1)}_{s} 
(k r)/2 $; then,
\begin{eqnarray}
G_{l+\nu} 
\left(
\left. 
r'', r' ;  
E
\right| 
\lambda; a
\right)
& = &
- \frac{2M}{\hbar^{2}} 
\,
\frac{\pi i}{2}
\, 
\sqrt{r'r''}
\;
\frac{ H^{(1)}_{i\Theta_{l}} (k r_{>})}{ H^{(1)}_{i\Theta_{l}}(k a)}
\nonumber
\\
& \times &
\left[
H^{(1)}_{i\Theta_{l}}
(k a) J_{i\Theta_{l}}(k r_{<})
-
J_{i\Theta_{l}}(ka) 
H^{(1)}_{i\Theta_{l}}
(k r_{<})
\right]
\;  ,
\label{eq:GreenF_l_reg_scatt}
\end{eqnarray}
where $k =
\sqrt{2ME}/\hbar$;
from this, 
the S-matrix can be derived with the formulation
of Ref.~\citen{ger:80}.

Equations~(\ref{eq:GreenF_l_reg})
and (\ref{eq:GreenF_l_reg_scatt})
 provide the
solution to our problem because the
energy Green's functions summarize all the 
physical information about bound states and scattering.
Indeed, Eq.~(\ref{eq:GreenF_l_reg}) is identical to the 
corresponding formula recently
found in Ref.~\citen{cam:pi_bs}.
This completes the proof of the equivalence between the
perturbative method of Ref.~\citen{cam:pi_bs}
and the nonperturbative approach of this paper.

\section{Renormalization, Anomalies, and Conclusions}
\label{sec:conclusions}
\noindent
The analysis that follows from Eq.~(\ref{eq:GreenF_l_reg}) 
is identical to that of Ref.~\citen{cam:pi_bs},
which we summarize here for completeness.
The bound-state sector of the theory
is determined by the poles of Eq.~(\ref{eq:GreenF_l_reg}),
which provide the implicit solutions of the equation $K_{s_{l}} (\kappa a) = 0$.
This equation provides no bound states in the subcritical regime,
but the imaginary index
$s_{l} =i \Theta_{l}$
changes the nature of the Bessel function and  leads to 
bound states in the supercritical regime. Then, 
as $a$ is just a real-space regulator, attention needs to be
focused only upon the limiting behavior for small $a$; 
consequently,
the energy levels become
$
E_{ n_{r} l }
=
-
\left( 2 \, e^{-\gamma}/a \right)^{2}
\,
\exp 
\left( - 2 \pi n_{r}/\Theta_{l} \right)
$
(with $n=n_{r}$ being the radial quantum number).
This result, properly renormalized
[with $\Theta_{l}=\Theta_{l} (a)$
in the limit $a \rightarrow 0$],
reproduces the known solution of the path-integral 
approach\cite{cam:pi_bs} and of the Hamiltonian operator
approach.\cite{gup:93,cam:isp,cam:dtII}

A remarkable property of this solution is the emergence of an arbitrary dimensional scale, 
within a theory that is devoid of dimensional
scales at the level of the Lagrangian---the phenomenon of 
dimensional transmutation.\cite{col:73}

An even more illuminating approach to this scale problem is provided by a general 
analysis of symmetry of the action (but not necessarily the Lagrangian) 
 under time reparametrizations.\cite{jac:72,alf:76,jac:80,jac:90}
 Invariance
of the action leads to the selection of the SO(2,1) conformal group, 
as for the inverse square potential,\cite{alf:76} the magnetic
monopole,\cite{jac:80} and the magnetic vortex.\cite{jac:90}
The corresponding generators are:
\begin{romanlist}
\item
 the Hamiltonian $H$,
associated with time translations $t
\rightarrow t - \alpha $;

\item
 the dilation generator
\begin{equation}
D= t H - \frac{1}{4} \left( {\bf r} \cdot {\bf p} + {\bf p} \cdot
{\bf r}
 \right)
 \;  ,
\end{equation}
associated with a scale transformation $t \rightarrow \tau t; 
{\bf r} \rightarrow \tau^{1/2} 
{\bf r}
$;

\item
 the
special conformal generator
\begin{equation}
K= H t^{2} - \frac{1}{2} ({\bf p} \cdot {\bf r} + {\bf r} \cdot
{\bf p} ) \,
 t + \frac{1}{2} m
r^{2}
 \;  ,
\end{equation}
associated with the special conformal transformation with respect to time
$1/t
\rightarrow 1/t + \alpha$.
\end{romanlist}
 These three generators satisfy the commutator relations:
$[D,H]= -i \hbar H$, $[D,K]= i \hbar K$, and $[H,K]=
2 i \hbar D$, which show
that they form an SO(2,1)
algebra. The same algebra is displayed by the two-dimensional delta-function 
interaction.\cite{jackiw:delta,cam:delta}
The anomalous version of this algebra as a consequence of
the breakdown of the classical SO(2,1) symmetry is currently being
investigated and will be reported elsewhere.

This quantum symmetry breaking of SO(2,1) invariance
is exhibited in Nature at the level of molecular physics, where it is realized in the point-dipole
approximation.\cite{cam:dipole}
In particular,  
the interaction between
an electron and a polar molecule 
can be reduced to a radial problem with 
an inverse square potential, for which the analysis of this paper applies.
This information is then transferred to the angular part of the problem,
which provides a 
critical dipole moment
for electron capture and formation of anions---a result 
of unusual relevance and which is
confirmed  by a large body of existing
experimental and numerical evidence.\cite{mea:84,des:94}
This is indeed the most remarkable example of a
quantum anomaly.\cite{cam:dipole}

In summary, we have completed a derivation
of the energy Green's function for the
regularized inverse square potential
(using a
regularization {\it \`{a} la\/} field theory)
in the (strong) supercritical 
regime.
This problem provides an example 
illustrating that the use of intrinsically nonperturbative
techniques (operator Green's function approach)
yields results identical to the corresponding ones derived
from the infinite summation of perturbation theory to all orders.

\nonumsection{Acknowledgments}
\noindent
We would like to thank
Profs.\ Roman Jackiw, 
Luis N. Epele, Huner Fanchiotti,
 and Carlos A. Garc\'{\i}a Canal
for discussions.
This research was supported in part by
an Advanced Research Grant from the Texas
Higher Education Coordinating Board 
and by the University of San Francisco Faculty Development Fund.

\nonumsection{References}


\begin{thebibliography}{000}
\bibitem{Wronskian}
The denominator 
$
p (r') \;  
W \left\{ 
u^{(<)}_{l,\nu} ,u^{(>)}_{l,\nu} 
\right\} 
(r') $
is guaranteed to be a constant due to the
self-adjoint nature of the given differential operator;
see Refs.~\citen{morse_feshbach} and~\citen{arfken}. This is a
particular case of Abel's formula for the Wronskian
(Ref.~\citen{stakgold}, p.\ 203).
For the specific problem discussed in this paper, 
that of Eq.~(\ref{eq:GF_radial}),
$W \left\{ 
u^{(<)}_{l,\nu} ,u^{(>)}_{l,\nu} 
\right\} 
(r') $
 itself is constant.
\end{thebibliography}
\end{document}